\documentstyle[prl,aps,multicol]{revtex}
\begin{document}
\draft

\title{ Growth Kinetics in a Phase Field Model with Continuous
Symmetry.}

\author{Umberto Marini Bettolo Marconi}
\address{Dipartimento di Matematica e Fisica, Universit\`a di Camerino,
         Via Madonna delle Carceri,I-62032 , Camerino, Italy}
\address{Unit\`a INFM di Camerino and Sez. INFN Perugia}
\author{Andrea Crisanti}
\address{Dipartimento di Fisica, Universit\`a di Roma ''La Sapienza''
         P.le A.Moro 2, I-00185 , Roma, Italy}
\address{Unit\`a INFM di Roma}

\date{\today}

\maketitle
\begin{abstract}
 We discuss the static and kinetic
 properties of a Ginzburg-Landau spherically
 symmetric $O(N)$ model recently introduced
 (Phys. Rev. Lett. {\bf 75}, 2176, (1995)) in order to
generalize the so called Phase field model of Langer.
The Hamiltonian contains two
$O(N)$ invariant fields  $\phi$ and $U$ bilinearly coupled.
The order parameter field
$\phi$ evolves according to a non conserved dynamics, whereas the diffusive
field $U$ follows a conserved dynamics. In the limit $N \to \infty$
we obtain an exact solution, which displays an
interesting kinetic behavior characterized by three different growth regimes.
In the early regime the system displays normal scaling
and the average domain size grows as $t^{1/2}$, in the intermediate
regime one observes a finite wavevector instability, which is related to the
Mullins-Sekerka instability; finally, in the late
stage the structure function has a multiscaling behavior, while the
domain size grows as $t^{1/4}$.

\end{abstract}


\begin{multicols}{2}
\section{Introduction}

When a system described by an order parameter, initially placed
into a high temperature single phase region
of its phase diagram, is brought to a point inside the coexistence curve
by a sudden change of temperature it becomes thermodynamically
unstable and  phase separates as a result of the existence of many competing
ground states.
 After the quench the system
can order kinetically through either nucleation or
spinodal decomposition. In the latter process a microscopic long-wavelength
fluctuation initially present is amplified and
determines the formation and evolution of various patterns characterized
by the presence of a universal length scale $L(t)$, associated with the
typical domain-size and separation among topological defects.
As the system orders, $L(t)$
grows in time in a power-law fashion $t^{1/z}$ and
the time-dependent structure factor $C(\bbox{k},t)$ displays dynamical scaling.

 A successful approach to the study of these phenomena is represented by the
time dependent Ginzburg-Landau  equation. Many years ago Hohenberg and
Halperin \cite{Hohenberg}
provided a useful classification of the various models,
which comprises a vast class of dynamic critical phenomena, in terms
of few parameters. Within their scheme two
models have received a great deal of attention: model A,
where a single field
evolves towards equilibrium with non conserved order parameter
dynamics (NCOP) and model B
where  the order parameter is conserved (COP) \cite{Bray}.

 The NCOP dynamics is aimed to describe an ordering process similar
to that of the Ising model with conventional Monte Carlo spin flip dynamics,
while COP accounts for the approach to equilibrium of an alloy.
In this case, one
observes that the system orders by growing droplets larger than
a critical size at the expenses of smaller droplets, while keeping
the total amount of material fixed.  The effect of the conservation law
is to slow down the phase separation process,
because the material has to be transported via diffusion through the system
before being added to a growing region.
 The value of the dynamical exponent $z$ is
$z=2$ for NCOP, whereas for conserved dynamics is
$z=3$ for a scalar order parameter and $z=4$ for
vector order parameter, indicating that conservation laws
play an important role in the dynamical process.

 While COP and NCOP dynamics has been widely studied,
the Phase field model \cite{Langer},  which describes
the coupling of a NCOP system with a diffusive COP field
such as temperature or concentration,
seems not to have been completely explored,
in spite of the fact that it displays
a variety of interesting peculiar features.
Only to mention the most striking of these,
we recall that the Phase field model
accounts for the regularity of the shapes observed during the
growth of crystals into a supercooled melt. According to the Phase field model
a planar solid front growing in the supercooled liquid undergoes the
so called Mullins-Sekerka instability. This phenomenon can be understood
as follows. The latent heat, released when the liquid freezes, is diffused into
the colder liquid and thus promotes the freezing of more material.
 The larger is the temperature gradient the faster is the
advancement of the front.
Now, imagine to slightly perturb the isothermal flat solid-liquid interface
in a slowly varying
fashion. As a result of the deformation,
the temperature gradient
will be larger on the bulges of such a boundary and so the heat flux.
This fact makes a solid tip to grow faster than a flat portion of interface
and provides a mechanism by which a perturbation of finite
wavelength is amplified, as it was discovered by Mullins and Sekerka
\cite{Pelce,Mullins}.

In a recent letter \cite{Noi}  we have introduced an $N$-component
version of the Phase field model in order to study the
evolution of a non conserved order parameter $\bbox{\phi}$ bilinearly
coupled to a conserved field $\bbox{U}$. In the limit
$N\to\infty$ we were able to
obtain some analytical results on the non-equilibrium relaxation
behaviour of the model. Here we expand these results.

The model, inspired by the  model C \cite{Hohenberg},
displays several interesting features: the evolution of the vector field
$\bbox{\phi}$ is non conserved in the early regime, then after a crossover time
it develops an instability at finite wavelength due to the coupling with
the conserved field $\bbox{U}$. In the very late regime the COP
behaviour becomes eventually dominant and $\bbox{\phi}$ shows a genuine
COP evolution, including multiscaling.
A similar mechanism
was reported by Somoza and Sagui \cite{Somoza} in a
numerical study of the  model C
where they observed that notwithstanding the non conserved field evolves faster
than the conserved field. For late times the growth is
driven by diffusion of the conserved variable and the order
parameter becomes slaved by the diffusive field.

The paper is organized as follows. In section \ref{sec:motivation} we
give a brief physical motivation of the model. The equation defining the
model are given in section \ref{sec:model}, and its equilibrium and dynamical
properties are discussed in section \ref{sec:equilibrium} and section
\ref{sec:dynamics}, respectively. Finally section \ref{sec:conclusions}
contains a brief summary of the results and discussions.

\section{Motivation of the Model}
\label{sec:motivation}
 The model we discuss in the present paper belongs to the family of the
$O(N)$ spherical models and it has been introduced \cite{Noi} with the aim of
studying exactly
the coupling of a NCOP field with a diffusive COP field. The $O(N)$
generalization proves to be fruitful because,
while retaining the salient features
of the phenomena occurring during the diffusion limited growth
it allows for some analytical results in the limit $N\to\infty$.

Historically, models containing couplings quadratic
with respect to the material NCOP field
$\phi$ and linear with respect to the COP  diffusive thermal field
$u$ were introduced as early as the seventies in the framework
of the dynamical critical phenomena and named  models C.
 Few years later, Langer, in order to study first order phase transitions
accompanied by latent heat of fusion,
put forward the so called Phase field model \cite{Langer},
in which the coupling was
assumed to be bilinear with respect to the two fields.

The material is characterized by an order parameter
$\phi$ which assumes
a positive value in the solid phase and a negative value in the
liquid phase. The local temperature of the system is treated as an
additional dynamical field obeying a heat diffusion equation in
the presence of sources represented by the amount of material changing phase.
 The solidification takes place adiabatically so that
no heat can flow to the outside.
 One defines the  dimensionless temperature field as
$u({\bbox x},t)=c_p(T({\bbox x},t)-T_m)/L$, where $L$ is the latent heat
of fusion per mole, $c_p$ the specific molar heat at constant pressure
and $T_m$ is the bulk melting temperature.
 The spatial average of $u$ at the initial time is the so called undercooling
parameter $\Delta$ and is a negative quantity.

 The thermal field diffuses according to the modified Fourier equation:
\begin{equation}
\label{eq:model0}
\frac{\partial u(\bbox {x},t)}{\partial t}
=D\nabla^2 u(\bbox {x},t)
+\frac{\partial\phi(\bbox{x},t)}{\partial t}
\end{equation}
where $D$ is the thermal diffusivity and the last term on the
r.h.s. is the amount of material which crystallizes
per unit time and thus proportional to the heat released
during the first order transition.
Both $c_p$ and $D$ are assumed to be equal in the two phases.
 The evolution of the order parameter is determined by the
non linear time-dependent equation of the Ginzburg-Landau type
\cite{Penrose,Wheeler,Kupferman}:
\begin{eqnarray}
   \frac{\partial \phi(\bbox{ x},t)}{\partial t} &=&
               - \Gamma\frac{\delta}{\delta\phi(\bbox{ x},t)}\, F[\phi,u]
 \nonumber\\
              &=& - \Gamma [-\nabla^2 \phi+ r \phi +g \phi^3 +\alpha u]
\label{eq:cap2}
\end{eqnarray}
To describe a two phase system
the form of $F$ is chosen to be a double
well for $r<0$. The coupling to the thermal
field $u$ can create an unbalance, in such a way that for negative values
of $\alpha u$ the liquid phase ($\phi \sim -\sqrt{-r/g}$)
is metastable with respect to the solid ($\phi \sim +\sqrt{-r/g})$.
In the absence of coupling  to the temperature field, i.e. $\alpha=0$,
eq. (\ref{eq:cap2}) represents the familiar  Cahn-Allen
equation, also called  Model A.

 The process contains two stages: during the first stage the solid grows
at the expenses of the liquid, while in the second stage the total amount
of solid is nearly constant and the growth is limited by diffusion
of the thermal field.

Interestingly, the two dynamical equations (\ref{eq:model0})
 and (\ref{eq:cap2})
can be derived from a Lyapounov functional ${\cal F}$, which
plays the role of the time dependent Ginzburg-Landau potential in the
present problem. If one performs the transformation
$U=u-\phi$ and eliminates $u$ in favor of the new field
$U$ one can write eqs. (\ref{eq:model0})-(\ref{eq:cap2}) as
\begin{eqnarray}
\frac{\partial \phi(\bbox{x},t)}{\partial t} &=&
-\Gamma\left.\frac{\delta {\cal F}}{\delta \phi({\bbox x},t)}
       \right|_{U}
\label{eq:quattro}
  \\
\frac{\partial U(\bbox{x},t)}{\partial t} &=& \frac{D}{\alpha}\,
               \nabla^2\left.\frac{\delta {\cal F}}{\delta U(\bbox{x},t)}
                       \right|_{\phi}
\label{eq:cinque}
\end{eqnarray}
with the Lyapounov functional
\begin{eqnarray}
\label{eq:tre}
{\cal F}[\phi,U]=\int d^dx\, \Bigl[\frac{1}{2}(\nabla \phi)^2&+&
\frac{r}{2} \phi^2+\frac{g}{4} \phi^4 \nonumber \\
&+&\frac{\alpha}{2} (U+\phi)^2\Bigr].
\end{eqnarray}
The functional ${\cal F}$ has
two equal minima when the temperature
field vanishes, i.e., for $U=-\phi$
and generates a complex dynamical behavior which
has been the object of some studies. However, its global properties
are not so well known. This fact, lead us to formulate an $O(N)$
invariant vectorial generalization of the above model. This kind of models,
in fact, lend themselves to nearly analytical solutions
thus providing
useful insights on the properties of the scalar order parameter solutions.

\section{The $O(N)$ model}
\label{sec:model}
We shall consider a system described by two coupled $N$-component vector fields
$\bbox{\phi} =(\phi_1({\bbox x},t),...,\phi_N({\bbox x},t))$ and
$\bbox{U}=(U_1({\bbox x},t),...,U_N({\bbox x},t))$,
whose Hamiltonian can be represented by \cite{Gunton,Wheeler}:
\begin{equation}
\label{eq:model}
\begin{array}{ll}
   H[\bbox{\phi}(\bbox{ x}),&\bbox{U}(\bbox{x},t)] = \displaystyle\int d^d x
 \left[\frac{1}{2}(\nabla \bbox{\phi})^2\right. \nonumber \\ \\
      &\displaystyle +\frac{r}{2} \bbox{\phi}^2
      \left.+\frac{g}{4N}( \bbox{\phi}^2)^2
      +\frac{w}{2} \bbox{U}^2 + \mu  \bbox {U} \bbox{\phi}\right]
\end{array}
\end{equation}
where $r$ and $g$, with $g>0$ and $w>0$, are the standard quadratic
and quartic couplings of the Ginzburg-Landau model
and the last term represent a bilinear coupling between the
field $\bbox{\phi}$ and $\bbox{U}$.
The first three terms in
eq. (\ref{eq:model}) constitute the familiar Ginzburg-Landau-Wilson
Hamiltonian describing an $O(N)$ $\phi^4$ model, whereas the last
two terms represent the interaction between the order parameter field
and an external fluctuating field $\bbox{U}$.

We assume that the field $\bbox{\phi}$ evolves according to NCOP
dynamics:
\begin{equation}
\label{eq:Lang1}
   \frac{\partial \phi_{\alpha}(\bbox{ x},t)}{\partial t} =
               - \Gamma_{\phi}\frac{\delta}{\delta
\phi_{\alpha}(\bbox{ x},t)}\, H[\bbox{\phi}, \bbox{U}]
               + \eta_{\alpha}(\bbox{x},t)
\end{equation}
whereas the field $\bbox{U}$ is conserved and relaxes according to:
\begin{equation}
\label{eq:Lang2}
   \frac{\partial U_{\alpha}(\bbox{ x},t)}{\partial t} =
               \Gamma_U\nabla^2 \frac{\delta}{\delta
U_{\alpha}(\bbox{ x},t)}\, H[\bbox{\phi},\bbox{U}]
               + \xi_{\alpha}(\bbox{x},t)
\end{equation}
The noises appearing on the right hand sides of eqs. (\ref{eq:Lang1})-
(\ref{eq:Lang2}) have zero average and two-point correlations:
\begin{eqnarray}
\langle\eta_{\alpha}(\bbox{x},t)&\,& \eta_{\beta}(\bbox{x'},t)\rangle =
\nonumber \\
 &\,&2\,T_f\,\Gamma_{\phi}\,\delta_{\alpha,\beta}\, \delta(\bbox{x-x'})\,
\delta(t-t')
\label{eq:noise1}
   \\
\langle\xi_{\alpha}(\bbox{x},t)&\,& \xi_{\beta}(\bbox{x'},t)\rangle =
\nonumber \\
 &\,&-2\,T_f\,\Gamma_{U}\,\delta_{\alpha,\beta}\,\nabla^2 \delta(\bbox{x-x'})\,
\delta(t-t')
\label{eq:noise2}
   \\
\langle\eta_{\alpha}(\bbox{x},t)&\,&\xi_{\beta}(\bbox{x'},t)\rangle = 0
\label{eq:noise3}
\end{eqnarray}
where
$T_f$ is the temperature of the
final equilibrium state whereas $\Gamma_U$ and $\Gamma_{\phi}$ are the
kinetic coefficients.

Introducing the Fourier components of the fields one can write the evolution
equation as,
\begin{eqnarray}
 \frac{\partial\phi_{\alpha} (\bbox{k},t)}{\partial t} &=&
       F_{\phi}^{\alpha}(\bbox{k}) + \eta_{\alpha}(\bbox{k},t)
 \label{eq:phi1}
     \\
 \frac{\partial U_{\alpha} (\bbox{k},t)}{\partial t} &=&
       F_{U}^{\alpha}(\bbox{k}) + \xi_{\alpha}(\bbox{k},t)
\label{eq:U1}
\end{eqnarray}
where $F_{\phi,U}$ are the Fourier transforms of the
first term on the r.h.s of eqs.(\ref{eq:Lang1}) and
(\ref{eq:Lang2}).

In the limit $N\to\infty$  the cubic term entering into $F_{\phi}$ can be
decoupled and we have:
\begin{eqnarray}
   F_{\phi}^{\alpha}(\bbox{k}) =
                   M_{\phi\phi}(k,t)\, \phi_{\alpha}(\bbox{k},t)
                 + M_{\phi U}(k,t)\,      U_{\alpha}(\bbox{k},t)
\label{eq:phi}
   \\
   F_{U}^{\alpha}(\bbox{k}) =
                   M_{U \phi}(k,t)\, \phi_{\alpha}(\bbox{k},t)
                 + M_{U U}(k,t)\,       U_{\alpha}(\bbox{k},t)
\label{eq:U}
\end{eqnarray}
where the matrix elements are given by
\begin{equation}
\label{eq:matrix}
\begin{array}{ll}
M_{\phi\phi}(k,t) =& -\Gamma_{\phi}[k^2+r+g S(t)], \\
M_{\phi U}(k,t)=& -\Gamma_{\phi} \mu \\
M_{U \phi}(k,t)=& -\Gamma_{U} \mu k^2, \\
M_{U U}(k,t) =&-\Gamma_U w k^2.
\end{array}
\end{equation}
The quantity $S(t)$ is the integrated ${\bf \phi}$-structure function
\begin{eqnarray}
  S(t) &=& \frac{1}{N} \sum_{\alpha}^N\langle\phi_{\alpha}(\bbox{x},t)\,
                                              \phi_{\alpha}(\bbox{x},t)\rangle
 \nonumber\\
       &=& \int\,\frac{d^d k}{(2 \pi)^d}\, \langle\phi_{\alpha}(\bbox{k},t)\,
                                              \phi_{\alpha}(-\bbox{k},t)\rangle
\label{eq:St}
\end{eqnarray}
and the integral contains a phenomenological momentum cutoff $\Lambda$.
The average
is over the external noises $\bbox{\eta}$ and $\bbox{\xi}$ and initial
conditions.

To study the behaviour at finite temperature $T_f$ it is useful to introduce
the equations of motion for the three
equal-time real space connected correlation functions
 $C_{\phi\phi}(r,t)=\langle\phi_{\alpha}(R+r,t) \phi_{\alpha}(R,t)\rangle$
 $C_{\phi U}(r,t)=\langle\phi_{\alpha}(R+r,t) U_{\alpha}(R,t)\rangle$ and
 $C_{UU}(r,t)=\langle U_{\alpha}(R+r,t) U_{\alpha}(R,t)\rangle$,
whose Fourier transforms are the structure functions. These correlations are
independent of the index $\alpha$ due to the internal symmetry.
In the $N\to\infty$ limit the structure functions
evolve according to the following set of equations:
\begin{eqnarray}
\label{eq:c11}
\frac{1}{2}\,\frac{\partial}{\partial t}\, C_{\phi\phi}(k,t) &=&
M_{\phi\phi}(k,t) C_{\phi\phi}(k,t) \nonumber \\
&+&M_{\phi U}(k,t) C_{\phi U}(k,t) +\Gamma_{\phi}T_f
\end{eqnarray}
\begin{eqnarray}
\label{eq:c12}
\frac{\partial}{\partial t}\, C_{\phi U}(k,t)&=&
M_{U\phi}(k,t) C_{\phi \phi}(k,t) \nonumber \\
&+&(M_{U U}(k,t)
+M_{\phi\phi}(k,t) )C_{\phi U}(k,t) \nonumber \\
&+&M_{\phi U}(k,t) C_{U U}(k,t)
\end{eqnarray}
\begin{eqnarray}
\label{eq:c22}
\frac{1}{2}\,\frac{\partial}{\partial t}\, C_{U U}(k,t)&=&
M_{U\phi}(k,t) C_{\phi U}(k,t) \nonumber \\
&+&M_{U U}(k,t) C_{U U}(k,t) + \Gamma_{U}T_f k^2.
\end{eqnarray}

In what follows we shall be more interested
into the behaviour of the field $\bbox{\phi}$, since it is the relevant order
parameter of the system.

\section{Equilibrium properties}
\label{sec:equilibrium}
In this section we investigate the equilibrium properties of the model
(\ref{eq:phi1}) and (\ref{eq:U1}).
It can be shown that the random process characterized by
the Langevin equations (\ref{eq:noise1})-(\ref{eq:U1})
obeys detailed balance since the following
``potential conditions'', analogous of the Onsager relations,
are fulfilled \cite{Gra73}:
\begin{eqnarray}
  \frac{\delta}{\delta \phi_{\alpha}(\bbox{k})}\,
                              F_{\phi}^{\beta}(-\bbox{k}') &=&
  \frac{\delta}{\delta \phi_{\beta}(\bbox{k}')}\, F_{\phi}^{\alpha}(-\bbox{k}),
\label{eq:grah1}
   \\
  \frac{\delta}{\delta \phi_{\alpha}(\bbox{k})}\, F_U^{\beta}(-\bbox{k}') &=&
  \frac{\Gamma_U\,k'^{\,2}}{\Gamma_{\phi}}\,
  \frac{\delta}{\delta U_{\beta}(\bbox{k}')}\, F_{\phi}^{\alpha}(-\bbox{k}),
\label{eq:grah2}
   \\
  \frac{\delta}{\delta U_{\alpha}(\bbox{k})}\, F_U^{\beta}(-\bbox{k}') &=&
  \frac{k'^{\,2}}{k^2}\,
  \frac{\delta}{\delta U_{\beta}(\bbox{k}')}\, F_U^{\alpha}(-\bbox{k}).
\label{eq:grah3}
\end{eqnarray}
If detailed balance holds, the stationary probability
density reads
\begin{equation}
 \label{eq:probst}
 P_{\rm st}[\bbox{\phi},\bbox{U}] =
    {\cal N} \exp\left(-\frac{1}{T_f}\,H[\bbox{\phi},\bbox{U}]\right)
\end{equation}
where ${\cal N}$ is a normalization constant.

The equilibrium probability density is quadratic in the field $\bbox{U}$,
therefore
as far as the static properties of $\bbox{\phi}$ are involved, the field
$\bbox{U}$ can be traced out. One is then left with an effective Hamiltonian
for the field $\bbox{\phi}$:
\begin{equation}
\label{eq:renorm1}
   H_{\rm eff}[\bbox{\phi}] = \int d^d x
       \left[  \frac{1}{2}(\nabla \bbox{\phi})^2
             + \frac{r_{\rm eff}}{2} \bbox{\phi}^2
             + \frac{g}{4N}( \bbox{\phi}^2)^2 \right]
\end{equation}
where $r_{\rm eff}=r-\mu^2/w$ is the ``renormalized mass''.
The importance of the field $\bbox{U}$ can be fully appreciated only in the
non-equilibrium dynamics of the system, as will be discussed in the next
section.

Before considering the
dynamics, we briefly discuss the static properties of this
model. Unlike the case where the $\bbox{U}$ is quenched
\cite{Depasquale} the system displays for space dimensions $d>2$
an order-disorder transition when $T_f$ is
lower than the critical temperature $T_c$.
In order to locate the critical surface $T_f=T_c(r,g,\mu,w)$ one considers the
long-range behaviour of the structure functions $C_{\phi\phi}(k)$, which can be
computed from (\ref{eq:renorm1}). The fourth order term makes the calculation
difficult for finite $N$. However for $N \to \infty$ we can use the Hartree
approximation, exact in this limit, and we readily obtain for $T>T_c$:
\begin{eqnarray}
 C_{\phi\phi}(k)&=& \frac{T_f}{k^2+r+gS_{\infty}-\mu^2/w}.
\label{eq:missing}
  \\
 S_{\infty}&=& \int_{|\bbox{k}|<\Lambda}\, \frac{d^d k}{(2 \pi)^d}\,
C_{\phi\phi}(k).
\label{eq:missing1}
\end{eqnarray}
For $T_f \leq T_c$ the structure function diverges at small $k$
because the full mass term $r+gS_{\infty}-\mu^2/w$ vanishes,
signaling the appearance of the ordered phase.
In fact, the model for $r<\mu^2/w$ and $g>0$ displays a high temperature
paramagnetic phase and a low temperature ordered phase.
The critical temperature is given by the usual form of the
$\phi^4$ theory
$T_c=(\mu^2/w-r)(d-2)/(g \Lambda^{d-2} K_d)$ with
$1/K_d=2 \pi^{d/2}\Gamma(d/2)$ where $\Gamma(x)$ is the Gamma
function.

For temperatures $T_f$ below $T_c$ there exists a non vanishing
order parameter $M=\langle\phi_{1}\rangle$,
which can be assumed to be directed along the
$\alpha=1$ direction without
loss of generality.
The $(N-1)$ components of the correlation function orthogonal to the
order parameter direction diverge at small $k$, reflecting the
existence of Nambu-Goldstone modes, i.e, excitations of vanishing energy
cost in the long wavelength limit.
The real space two point correlation function takes the form
\begin{equation}
\label{eq:uno}
\langle\phi_{\alpha}(r)\,\phi_{\alpha}(r)\rangle=M^2 \delta_{\alpha 1}
                +S_{\infty}(t)
\end{equation}
\begin{equation}
\label{eq:op}
 M^2= - \frac{1}{g}\,\left(r-\frac{\mu^2}{w}\right)\,
\left(1 - \frac{T_f}{T_c}\right)
\end{equation}
where $S_{\infty}$ defined in equation (\ref{eq:missing1}) comes from the
transverse components only.

The other equilibrium correlation functions can also be obtained from the
stationary equilibrium distribution and read for $T>T_c$:
\begin{eqnarray}
 C_{UU}(k)&=& \frac{T_f}{w-\mu^2\,(k^2+r+gS_{\infty})^{-1}}
\label{eq:UUst}
  \\
 C_{\phi U}(k)&=& - \left(\frac{\mu}{w}\right)\,
            \frac{T_f }{k^2+r+gS_{\infty}-\mu^2/w}
\label{eq:phiUst}
\end{eqnarray}
Note that both  $C_{\phi U}(k)$ and  $C_{UU}(k)$ are singular for $k\to 0$.

We conclude by noting that to obtain the static structure functions from the
dynamical equations one has to supplement the requirement that
the right hand sides of eqs.(\ref{eq:c11})-(\ref{eq:c22}) vanish with
the following stronger condition:
\begin{eqnarray}
\label{eq:equ1}
\lim_{t \to \infty}\,\Bigl[ M_{U\phi}(k,t)&\,& C_{\phi \phi}(k,t)
\nonumber \\
                         &\,&+ M_{U U}(k,t)\, C_{\phi U}(k,t) \Bigr] =0,
\end{eqnarray}
\begin{eqnarray}
\label{eq:equi2}
\lim_{t \to \infty}\,\Bigl[ M_{\phi\phi}(k,t)&\,& C_{\phi U}(k,t)
\nonumber \\
                          &\,&+ M_{\phi U}(k,t)\, C_{U U}(k,t) \Bigr] =0
\end{eqnarray}
to ensure that the equilibrium properties of the model are independent
on the kinetic coefficients $\Gamma_{\phi}$ and $\Gamma_{U}$.
The conditions (\ref{eq:equ1})-(\ref{eq:equi2}) can also be deduced from the
the equilibrium properties of the model (see Appendix).

\section{Dynamical properties}
\label{sec:dynamics}

Since the behavior at $T_f=0$ is representative of the entire
dynamics in the ordered phase when $T_f < T_c$, we shall neglect the noise
terms in the following analysis \cite{Bray}.

For general initial conditions the two fields are not in equilibrium,
and we may expect that the relaxation of $\bbox{\phi}$ is only slightly
modified by the external field $\bbox{U}$. Since the
dynamics of $\bbox{U}$ is sufficiently slow compared with that of
$\bbox{\phi}$, the
presence of $\bbox{U}$ does not modify qualitatively the NCOP behaviour of
$\bbox{\phi}$.
In particular, the size of the domains of $\bbox{\phi}$ should grow with
a characteristic length $L(t)\sim t^{1/2}$, while the maximum of
the structure factor is located at $k=0$ and should increase in time with
the power $t^{d/2}$.

This kind of behavior persists until the domain size reaches
the typical length associated with the field $\bbox{U}$,
and given by the maximum of the
structure function of $\bbox{U}$. At this stage  the dynamics of
$\bbox{\phi}$ slows down because the coupling with the conserved field
$\bbox{U}$ introduces an additional
constraint on the dynamics of $\bbox{\phi}$. For
longer times the two fields equilibrate and the COP behaviour eventually
becomes dominant.

A simple analysis of the equation of motion for $N\to \infty$
gives the scaling of
the crossover time with the coupling constant $\mu$. Indeed it is simple
to see that making the rescalings
\begin{equation}
\label{eq:rescaling}
 \begin{array}{llll}
	t\,\mu^2    & \to t; &\phantom{xxx}
	k/\mu       & \to k; \\
	U/\mu       & \to U; &\phantom{xxx}
	r/\mu       & \to r; \\
        g/\mu^{d-1} & \to g; &\phantom{xxx}
	\Lambda/\mu & \to \Lambda;
 \end{array}
\end{equation}
the parameter $\mu$ disappears from equations of motion for
$\phi\equiv\phi_{\alpha}$ and $U\equiv U_{\alpha}$:
\begin{eqnarray}
 \frac{\partial}{\partial t}\, \phi(k,t) &=&
                    -\Gamma_{\phi}[k^2+r+g S(t)]\, \phi(k,t) \nonumber \\
                & & \phantom{xxxxxx}-\Gamma_{\phi} \mu\, U(k,t)
 \label{eq:phiT}
   \\
 \frac{\partial}{\partial t}\, U(k,t) &=& -\Gamma_{U} \mu k^2\, \phi(k,t)
                                        -\Gamma_U w k^2\, U(k,t).
\label{eq:UT}
\end{eqnarray}
As a consequence
the crossover time scales as $1/\mu^2$. From this analysis it follows that
if the dynamics of $U$ is sufficiently slow then for
$1 \ll t \ll 1/\mu^2$ the field $\phi$ exhibits a NCOP behaviour while for
$t \gg 1/\mu^2$ a COP behaviour.
If the dynamics of $U$ becomes too fast the first NCOP behaviour
shrinks and becomes hardly observable.

This scenario can be confirmed by solving the equation of motion
(\ref{eq:phiT}) and (\ref{eq:UT})
in a quasilinear approximation. To this end we assume that
\begin{equation}
\label{eq:R}
 R(t) = r + g\,S(t)
\end{equation}
is slowly varying in time, so that it can be considered constant over
successive
intervals of time. In other words, we make a piecewise linearization of the
equation of motion along the trajectory.
In spite of that, the approximation is
sufficient to identify the different regimes of the relaxation process.

If we neglect the time dependence of
$R(t)$ and assume it to be nearly constant eqs. (\ref{eq:phiT}) and
(\ref{eq:UT})
become a linear system whose solution has the form:
\begin{equation}
\label{eq:phib}
\begin{array}{ll}
  \phi(k,t) &=   c_{\phi}^{+}(k)\, e^{\omega_{+}(k)t}
               + c_{\phi}^{-}(k)\, e^{\omega_{-}(k)t} \\
  U(k,t)   &=   c_{U}^{+}(k)\, e^{\omega_{+}(k)t}
              + c_{U}^{-}(k)\, e^{\omega_{-}(k)t}
\end{array}
\end{equation}
where $\omega_{+}(k)$ and $\omega_{-}(k)$ are the eigenvalues of
the $M$ matrix,
\begin{eqnarray}
\label{eq:eigen}
   \omega_{\pm}(k)=\frac{1}{2}\,\Bigl[&-&\Gamma_{\phi}(k^2+R)-\Gamma_U w k^2
\nonumber \\
                   &\pm& \sqrt{[\Gamma_{\phi}(k^2+R)+\Gamma_U w k^2]^2+
                             4\,\Gamma_{\phi}\Gamma_U \mu^2 k^2}
                                \Bigr] \nonumber \\
& &
\end{eqnarray}
For time $t\gg 1$ the dynamical behavior of the solution is
determined by the larger eigenvalue $\omega_{+}(k)$.
For large  $k^2$, the eigenvalue $\omega_{+}(k)$ decreases proportionally to
$-k^2$ and hence large momenta are exponentially damped. Moreover
we see that $\omega_{+}(k)$ is a function of $k^2$ which
either has an extremum
at $k=0$ or a single maximum for $k\not= 0$, as one can
verify by inspecting  the small-$k$ behaviour of $\omega_{+}(k)$.
The behavior of $\omega{+}(k)$ is shown in Fig.~\ref{fig:dispersion}
for $t<\tau_f$ and $t>\tau_f$. The crossover time $\tau_f$ is defined as the
time
when the fastest growing mode moves from $k=0$ to $k\not= 0$.
Other definitions of $\tau_f$ are possible, e.g., the time when the
peak at $k\not=0$ becomes higher then the $k=0$ one. However, all definitions
lead to similar results.

Below the critical temperature $T_c$
and in the early stage of the ordering process the value
of $gS(t)$ is small compared with $r$, i.e. $R<0$, and the larger
eigenvalue is well approximated by:
\begin{equation}
\label{eq:dispersion1}
  \omega_{+}(k)= \Gamma_{\phi}|R|-
                 \left(\Gamma_{\phi}-\Gamma_{U}\frac{\mu^2}
{|R|}\right)\,k^2 + O(k^4)
\end{equation}
A brief calculation reveals that
\begin{eqnarray}
 c_{\phi}^{+} &=& \phi(k,0) + \frac{\mu}{R}\,U(k,0) + O(k^2)
\label{eq:coeffcp}
   \\
 c_{U}^{+} &=& \frac{\Gamma_U\mu}{\Gamma_{\phi}R}\,
               \left[\phi(k,0) + \frac{\mu}{R}\,U(k,0)\right]\,k^2 + O(k^4).
\label{eq:coeffcU}
\end{eqnarray}
Therefore, assuming $\phi(k,0) + (\mu/R)\,U(k,0) = O(1)$ for $k\to 0$,
the coefficient
$c_{\phi}^{+}(k)$ is finite for $k\to 0$ while
$c_U^{+}(k)$ vanishes, indicating that
the amplitudes of the longest wavelength components
of $U$ are decreased due to the conservation law.

As a consequence, below $T_c$ the structure factor $C_{\phi\phi}(k,t)$
develops a peak centered at $k=0$, growing in time as a power of $t$.
The structure functions $C_{\phi U}(k,t)$ and $C_{UU}(k,t)$
also develop a peak, centered at a finite value
of $k$, say $k_f$, as a result of the
competing effect between the $k$ dependence
of the exponential factor
$\exp [\omega_{+}(k)\,t]$ and the amplitude
$c_U^{+}(k)$, see eq (\ref{eq:coeffcU}).
 This mechanism selects a set of exponentially growing $U$-modes with
wavevectors in a certain range centered around $k_f$, whose dependence
on the coupling $\mu$ is shown in Fig.~\ref{fig:mu}. Such modes
represent an inhomogeneity of the $U$ field which in turn
affects the spatial properties of the $\phi$ subsystem.
One witnesses a strong feedback process between the two fields and the
outcome is the slaving of the NCOP dynamics of the field $\phi$
to the COP dynamics of the $U$-field.

The power law growth of $C_{\phi\phi}(k=0,t)$ can be extracted from
the quasilinear approximation by using (\ref{eq:dispersion1}). In the
early  regime
$R$ starts from a negative value and grows towards zero due to the growing of
$C_{\phi\phi}(k,t)$ for small $k$. This in turn implies that $S(t)$
tends to a finite value for increasing time. By imposing this condition, and
making use of (\ref{eq:dispersion1}) it follows that
$C_{\phi\phi}(k=0,t)\sim t^{d/2}$,
as in the pure NCOP, i.e. the longest
wavelength fluctuations grow faster. We note that while the quasilinear
approximation  leads to the correct scaling of
$C_{\phi\phi}(k=0,t)\sim t^{d/2}$ and of the domain size
$L(t)\sim t^{1/2}$, it gives for
$R(t)\sim \log(t)/t$ which reveals that the approximation is
slightly crude.

These results are valid for $R/\mu^2$ not too large, i.e., far from the
crossover region where $R$ changes sign.
Unlike the pure NCOP, where  $R(t)$ goes to zero,
after a characteristic time $\tau_f=O(1/\mu^2)$ the value of $|R|$
becomes $O(\mu^2)$ and the NCOP behaviour ends.
By inspection of eq.(\ref{eq:dispersion1}), we see that if
$|R|/\mu^2 < \Gamma_U/\Gamma_\phi$ the maximum of $\omega_{+}(k)$
moves away from $k=0$ and the system looses its NCOP behaviour.

This regime corresponds in our model to the instability which is
observed in systems where a non-conserved order parameter is coupled to a
conserved field.
We must stress that in order to observe the NCOP behaviour
the dynamics of $U$ needs to be sufficiently slow with respect to
the characteristic time $\tau_f\sim 1/\mu^2$, whose dependence on
$\mu$ is shown in Fig.~\ref{fig:tau}.

For times $t=O(1/\mu^2)$ the quantity $R$ changes sign becoming positive,
and finally for $t\to\infty$
tends to a finite value $\mu^2/w$ while the maximum of
$\omega_{+}(k)$ moves again towards vanishing wavevectors.
The dynamics is therefore
dominated in the regime $t\gg 1/\mu^2$ by long wavelength
fluctuations. We can then expand
$\omega_{+}(k)$ in powers of $k$ obtaining
\begin{equation}
\label{eq:dispersion2}
 \omega_{+}(k)=\Gamma_{U}\left( \frac{\mu^2}{r+gS}-w\right)k^2 - c_4 k^4
\end{equation}
where $c_4$ is a positive coefficient having a finite limit for
$\mu^2/R \to w$.
By imposing that $S(t)$ has a finite non zero, limit
for $t\to\infty$ and making use of
eq.(\ref{eq:dispersion2}) one obtains
that in this regime
\begin{equation}
\label{eq:COP}
C_{\phi\phi}(k,t) = \left[L(t)^2\,k_m(t)^{2-d}\right]^{\varphi(k/k_m(t))}
\end{equation}
where
\begin{equation}
\label{eq:COP1}
 L(t)\sim t^{1/4}, \qquad
 k_m(t) \sim \left(\frac{d}{4}\,\frac{\log t}{t}\right)^{1/4}
\end{equation}
a behaviour typical of COP dynamics \cite{Zann}. The multiscaling
function $\varphi(x)$ is
given by
\begin{equation}
\label{eq:multisc}
 \varphi(x) = 1 - (x^2 - 1)^2
\end{equation}
The  COP behaviour is also observed if one considers the structure functions
$C_{\phi U}(k,t)$ and $C_{UU}(k,t)$.
Such a multiscaling behavior follows from the competition of
two marginally distinct lengths, namely
the domain size $L(t)$ and $k_m^{-1}$.

Finally we note that the quasilinear approximation in this regime leads to
\begin{equation}
 \label{eq:rCOP}
  R(t) - \mu^2/w \sim \left( \frac{\log t}{t} \right)^{1/2}.
\end{equation}
 From equation (\ref{eq:missing}) we see that $R - \mu^2/w$ plays the role
of the mass term $r + g\,S(t)$ in pure COP dynamics \cite{Zann}, therefore
in spite of the fact that
the quasilinear approximation is quite crude, it gives
nevertheless the correct scaling behaviour of COP dynamics.

 The above theoretical predictions were checked
by integrating
numerically the system of equations (\ref{eq:c11})-(\ref{eq:c22}) by
the Euler method. The $k$ integrals were evaluated
by a Simpson rule discretizing the wavevectors in the interval $[0,\Lambda]$.
 Figure \ref{fig:fig4} displays the structure function $C_{\phi\phi}(k,t)$ for
various values of the time $t$. One clearly sees that in the early regime
the fastest growing modes are centered about $k=0$,
because long-wavelengths fluctuations
of the field $\phi$ increase more rapidly than shorter ones,
whereas for  $t>\tau_f$ a finite wavevector peak appears.
Moreover, the growth, in this late regime, has a conserved character
because its value at $k=0$ remains constant.
 The evolution of $C_{\phi U}(k,t)$
and $C_{UU}(k,t)$ is shown in Figs. \ref{fig:fig5} and \ref{fig:fig6}.
Initially the fields $U$ and $\phi$  evolve as if they were nearly independent
and the $C_{UU}$ correlations display the usual finite wavevector peak
of the conserved dynamics, whereas
the $\phi$ field evolves according to
a faster non conserved dynamics.
 The long
wavelength fluctuations of the $\bbox{U}$ field are hindered by
the conservation law and the presence of the $C_{\phi U}$ term
has only a small effect on the $C_{\phi\phi}$.
However, as the domain
size $L(t)$  reaches a critical value
and becomes comparable with  $\lambda_f$, the typical length
of the oscillations of the
diffusive field, the two fields strongly interact.
  Within this late regime the
dynamics becomes controlled by the conservation law induced  by the
$U$ field.
 In Fig. \ref{fig:fig7} we show the behavior of the height of the peak
of $C_{\phi\phi}$ versus time, where one clearly sees the crossover from
the early time behavior $t^{d/2}$ to the late stage slope $t^{d/4}$. In
the crossover regime due to the presence of a double peak the
maximum height decreases until the peak at $k=0$ disappears.

Finally we report the numerical result concerning
the multiscaling, observed in
the late regime. In Fig. \ref{fig:fig8} we display the shape function
$F(x)=k^{d}_m(t)C_{\phi\phi}(xk_m,t)$ as a function of $x$. In Fig.
\ref{fig:fig9}
we show the multiscaling function $\varphi(x)$ obtained from the best fit of
$C_{\phi\phi}(k,t)$ as a function of $L(t)^2\,k_m(t)^{2-d}$ for
fixed values of $x = k/k_m(t)$. Similar curves can
be extracted from the other
structure functions. We note that while the data follow quite well the
theoretical result (\ref{eq:multisc}) for $|x-1|$ not to large, they
display a large deviation
as $|x-1|$ increases. This is due by the terms neglected in
(\ref{eq:dispersion2}). We remark, however, that these
become less and less important
as $k_m$ decreases, as a consequence we expect that the range of values
of $|x-1|$ where there is a good agreement with (\ref{eq:multisc}) should
increase
with time. This is indeed observed by using data for increasing time in the
best fit. Roughly the range increases as $1/k_m(t)$.

Finally, we have explored, different types of conservation laws represented
by $\Gamma_U k^{\mu}$ with $0<\mu<2$. In all these cases the dynamics
selects for intermediate times a peak at finite values of th
wavevector $k$.

\section{Conclusions}
\label{sec:conclusions}
 To summarize
in this paper we have studied the evolution of an $N$-component
version of the Phase field model and shown that the coupling
between the massless transverse
modes and the diffusive field produce an instability at finite
wavelengths.

Our model, where the low-energy Nambu-Goldstone modes couple to
the diffusive modes
provides an interesting
new scenario and
we believe represents a paradigm for the
Mullins-Sekerka type of phenomena
where the soft modes are
represented by the capillary wave spectrum associated with the
solid-melt interface and the diffusive mode is the heat transport.
These two fields concur to destabilize the
solid-melt boundary in analogy with our findings.
On physical grounds, one expects this kind of instability to occur during
phase separation, because small droplets can dissipate
heat more efficiently and reach rapidly thermal equilibrium
due to the Gibbs-Thomson effect, whereas larger droplets try to
dissipate energy faster by creating bulges thus increasing
the curvature.
As the system cools down upon reaching equilibrium the
typical length of the bulges, $\lambda_f$, increases
and diverges together with the average  domain-size $L(t)$.

We have demonstrated that
the presence of $\bbox {U}$ induces non trivial
effects on the field $\bbox{\phi}$ because it acts
on a time scale longer than the noise field, characterized by
a short correlation time.

Finally, the $O(N)$ model analysed presents unusual
features since it displays
scaling behavior in the early regime ($t<\tau_f$) and multiscaling
\cite{Zannetti} in the late regime and constitutes an example of
multiscaling without COP, a phenomenon which at the best of our knowledge
was not observed before.

In summary the present model reveals an unexpectedly rich dynamical
behavior which reminds in many respects the solidification kinetics.

\section{Appendix: Equilibrium properties}
In the present appendix we shall outline the calculations of the
equilibrium properties of the model.
 The partition function associated with the Hamiltonian
\begin{equation}
Z[\{\bbox{h}(x)\}]= \int^{\infty}_{-\infty} \Pi_{\alpha=1}^N d\phi_{\alpha}
dU_{\alpha}\,
e^{-\beta H[{\bbox\phi},\bbox{U}]+\beta{\bbox h}{\bbox \phi}}
\label{eq:ZN}
\end{equation}
Where we have included an external field $\bbox{h}(\bbox{x})$ coupled linearly
to $\bbox{\phi}(\bbox{x})$ and $\beta=(k_B T_f)^{-1}$.
 The field $\bbox{U}(\bbox{x})$ can be traced out  and one finds a part from
uninteresting constants:
\begin{equation}
Z[\{\bbox{h}(x)\}]=\int^{\infty}_{-\infty}\Pi_{\alpha=1}^N d\phi_{\alpha}\,
e^{-\beta H_{\rm eff}[{\bbox\phi}]+\beta{\bbox h}{\bbox \phi}}
\label{eq:Zeff}
\end{equation}
where $H_{\rm eff}$ is defined by Eq. (\ref{eq:renorm1}).
In order to separate the macroscopic component $P$ of the field we
employ the following identity
\begin{equation}
1=N\int^{\infty}_{-\infty}
d P^2 \delta(NP^2-\sum_{\alpha}\phi_{\alpha}^2)
\label{eq:id}
\end{equation}
and rewrite $Z$ as
\begin{eqnarray}
Z= &N \int_{-\infty}^{\infty} dP^2\int_{-\infty}^{\infty}
\frac{d \lambda}{2 \pi} \int^{\infty}_{-\infty}\Pi_{\alpha=1}^N
d\phi_{\alpha}\, \nonumber \\
&\times e^{-\beta H_{\rm eff}
[\phi_{\alpha}]+\beta \bbox{ h \phi}+i \lambda(NP^2-\sum_{\alpha}
\phi_{\alpha}^2)}
\label{eq:ZN2}
\end{eqnarray}

\begin{eqnarray}
Z&=& N\int_{\infty}^{\infty} dP^2\int_{-\infty}^{\infty}
   \frac{d \lambda}{2 \pi}\,
   e^{-\beta N\left(\frac{r_{\rm eff}}{2} P^2+\frac{g}{4}
   P^4-i\frac{ \lambda}{\beta} P^2\right)} \nonumber\\
& &\times\int_{-\infty}^{\infty}\Pi_{\alpha=1}^N d\phi_{\alpha}
\nonumber \\
& &\times e^{-\beta/2 \sum_{\alpha}[-\int d^dx
                    \nabla\phi_{\alpha}\nabla\phi_{\alpha}
+2i \lambda/\beta \phi_{\alpha}^2]+\beta\sum_{\alpha} h_{\alpha}\phi_{\alpha}}
\label{eq:ZN3}
\end{eqnarray}
In the case of a uniform external field directed along the component 1
($h_1=h$), after
eliminating the ${\phi_{\alpha}}$ fields, $Z$ reads:

\begin{equation}
\label{eq:ZN4.1}
Z=N e^{\frac{N}{2}\ln(2\pi/\beta)}
\int_{-\infty}^{\infty} dP^2\int_{-\infty}^{\infty}
\frac{d \lambda}{2 \pi}\ e^{N\Omega}
\end{equation}
\begin{eqnarray}
\label{eq:ZN4.2}
\Omega =& -\beta \int d^dx \left[\frac{r_{\rm eff}}{2} P^2+\frac{g}{4}
        P^4-i\frac{ \lambda}{\beta} P^2\right] \nonumber \\
        & -\frac{1}{2}\,\int \frac{d^dk}{(2 \pi)^d}
        \ln (k^2+2 i \lambda/\beta)
        -i\frac{\beta^2 h^2}{4 \lambda}
\end{eqnarray}
In order to evaluate $Z$ we apply the saddle point estimate in
the limit $N\to \infty$ imposing the conditions
$(\partial Z / \partial \lambda)=0$ and
$(\partial Z / \partial P^2)=0$ which lead to the conditions:
\begin{equation}
\frac{2 i \lambda}{\beta}= r_{\rm eff} +g P^2
\label{eq:cond1}
\end{equation}
\begin{equation}
P^2=\int \frac{d^dk}{(2 \pi)^d}
\frac{1}{k^2+2 i \lambda/\beta}
-\frac{\beta^2 h^2}{4 \lambda^2}
\label{eq:cond2}
\end{equation}
Eliminating $\lambda$ with the help of eqs. (\ref{eq:cond1})-(\ref{eq:cond2})
we find:
\begin{eqnarray}
P^2&=&\frac{h^2 }{(r_{\rm eff}+g P^2)^2}+
\frac{1}{\beta}\int \frac{d^dk}{(2 \pi)^d}\, \frac{1}{k^2+r_{\rm eff}+gP^2}
\nonumber \\
&=& m^2 + S_{\infty}
\label{eq:cond2b}
\end{eqnarray}
 The last term equals $m^2$, the square of the average magnetization
per unit volume $ m =\frac{1}{V} d\ln  Z_N/dh$. By using
eq. (\ref{eq:cond2}) we find explicitly:
\begin{equation}
m=\frac{\beta}{2 i \lambda}=\frac{h }{r_{\rm eff}+g P^2}
\label{eq:cond3}
\end{equation}
The existence of a spontaneous magnetic phase
implies that in zero magnetic external field $m \neq 0$, i.e.
the following condition must
be fulfilled:
\begin{equation}
\lim_{h\to 0} \; [r_{\rm eff}+gP^2]=0
\label{eq:cond4}
\end{equation}
The equation of state reads
\begin{eqnarray}
\Bigl[ g T_f \int \frac{d^dk}{(2 \pi)^d}&\,&
        \frac{1}{k^2+r_{\rm eff} +gS_{\infty}
        +g m^2} \nonumber \\
&\,&+r_{\rm eff}+ gm^2 \Bigr]\, m = h
\label{eq:cond5}
\end{eqnarray}
where $S_{\infty}$ is given by:
\begin{equation}
S_{\infty}=
g T_f\int \frac{d^dk}{(2 \pi)^d}\,\frac{1}{k^2+r_{\rm eff} +gS_{\infty}
+g m^2}
\label{eq:cond6}
\end{equation}
In order to determine the critical temperature $T_c$ we require $m^2=0$
and $r_{\rm eff}+g S_{\infty}=0$.

\begin{equation}
T_c=(\mu^2/w-r)(d-2)/(g \Lambda^{d-2} K_d)
\label{eq:cond7}
\end{equation}
where
$1/K_d=2 \pi^{d/2}\Gamma(d/2)$ where $\Gamma(x)$ is the Gamma
function.

\narrowtext
\begin{figure}
\caption{Dispersion relation $\omega_{+}(k)$ in the early regime and in
the late regime.
         Both quantities are plotted in arbitrary units.
Notice the different long-wavelength behavior in the
two cases. The first is typical of NCOP dynamics, whereas the second
characterizes COP dynamics.}
\label{fig:dispersion}
\end{figure}

\begin{figure}
\caption{Dependence of the wavevector instability $k_f$ on the coupling
         parameter $\mu$ in log-log scale.
         Both quantities are plotted in arbitrary units.
         The squares are obtained from the numerical solution of
         eqs. (\protect\ref{eq:c11}) - (\protect\ref{eq:c22}) with
         $\Gamma_{\phi} = 1$, $\Gamma_U = 5$, $r = -0.5$, $g = 1$,
         $w = 0.05$ and
         $d=3$. The broken line has slope $1$.
         The value of $k_f$ is taken when the peak at $k_f$ becomes larger
         than the one at $k=0$. Other definitions, e.g., fastest growing mode,
         lead
         to a similar dependence as follows from the rescaling
         (\protect\ref{eq:rescaling}).
        }
\label{fig:mu}
\end{figure}

\begin{figure}
\caption{Dependence of the crossover time $\tau_f$ on the coupling
         parameter $\mu$ in log-log scale.
         Both quantities are plotted in arbitrary units.
         The broken line has
         slope $-2$. The squares are obtained from the numerical solution of
         eqs. (\protect\ref{eq:c11}) - (\protect\ref{eq:c22}) with
         $\Gamma_{\phi} = 1$, $\Gamma_U = 5$, $r = -0.5$, $g = 1$,
         $w = 0.05$ and
         $d=3$. The crossover time $\tau_f$ is defined as the time when the
         peak at $k\not= 0$ becomes dominant. Other definitions, e.g., fastest
         growing mode lead to similar dependence as follows from the rescaling
         (\protect\ref{eq:rescaling}).
        }
\label{fig:tau}
\end{figure}

\begin{figure}
\caption{Typical evolution of the structure function $C_{\phi\phi}(k,t)$
         for different times as shown in the figure.
         Both quantities are plotted in arbitrary units.
         Notice that the two
         largest times correspond to $t>\tau_f$ and display the characteristic
         COP peak.
         The data are from the numerical solution of
         eqs. (\protect\ref{eq:c11}) - (\protect\ref{eq:c22}) with
         $\Gamma_{\phi} = 1$, $\Gamma_U = 5$, $r = -0.5$, $g = 1$,
         $w = 0.05$ and
         $d=3$. }
\label{fig:fig4}
\end{figure}

\begin{figure}
\caption{Typical evolution of the structure function $C_{\phi U}(k,t)$
         The data are from the numerical solution of
         eqs. (\protect\ref{eq:c11}) - (\protect\ref{eq:c22}) with
         parameters as in Fig. \protect\ref{fig:fig4}.
         Both quantities are plotted in arbitrary units.
	 }
\label{fig:fig5}

\end{figure}
\begin{figure}
\caption{Time evolution of the structure function $C_{UU}(k,t)$
         for different times. Notice the conserved dynamics at all times.
         The data are from the numerical solution of
         eqs. (\protect\ref{eq:c11}) - (\protect\ref{eq:c22}) with
         parameters as in Fig. \protect\ref{fig:fig4}.
         Both quantities are plotted in arbitrary units.
         }
\label{fig:fig6}
\end{figure}

\begin{figure}
\caption{Height of the peak of the structure function
         $C_{\phi\phi}(k_{\rm m},t)$
         as a function of time.
         Both quantities are plotted in arbitrary units.
         The crossover from the NCOP behavior to the
         COP behavior is evident. The dashed line represents the $t^{d/2}$
         behavior
         whereas the dashed-dot line the $t^{d/4}$ behavior.
         The data are from the numerical solution of
         eqs. (\protect\ref{eq:c11}) - (\protect\ref{eq:c22}) with
         parameters as in Fig. \protect\ref{fig:fig4}.
         }
\label{fig:fig7}
\end{figure}

\begin{figure}
\caption{Shape function $F(x)$ of the
         $C_{\phi,\phi}(k,t)$ structure function
         in the late stage evolution. The absence of scaling
         is evident in figure.
         The data are from the numerical solution of
         eqs. (\protect\ref{eq:c11}) - (\protect\ref{eq:c22}) with
         parameters as in Fig. \protect\ref{fig:fig4}.
         }
\label{fig:fig8}
\end{figure}

\begin{figure}
\caption{Multiscaling exponent $\varphi(x)$ defined in the text.
         The data are from the numerical solution of
         eqs. (\protect\ref{eq:c11}) - (\protect\ref{eq:c22}) with
         $\Gamma_{\phi} = 1$, $\Gamma_U = 5$, $r = -0.5$, $g = 1$,
         $w = 0.05$ and
         $d=3$.
         The crosses are obtained from $C_{\phi\phi}$ while the triangles
         from
         $C_{UU}$. The full line is the theoretical prediction
         (\protect\ref{eq:multisc}).
         }
\label{fig:fig9}
\end{figure}

\end{multicols}
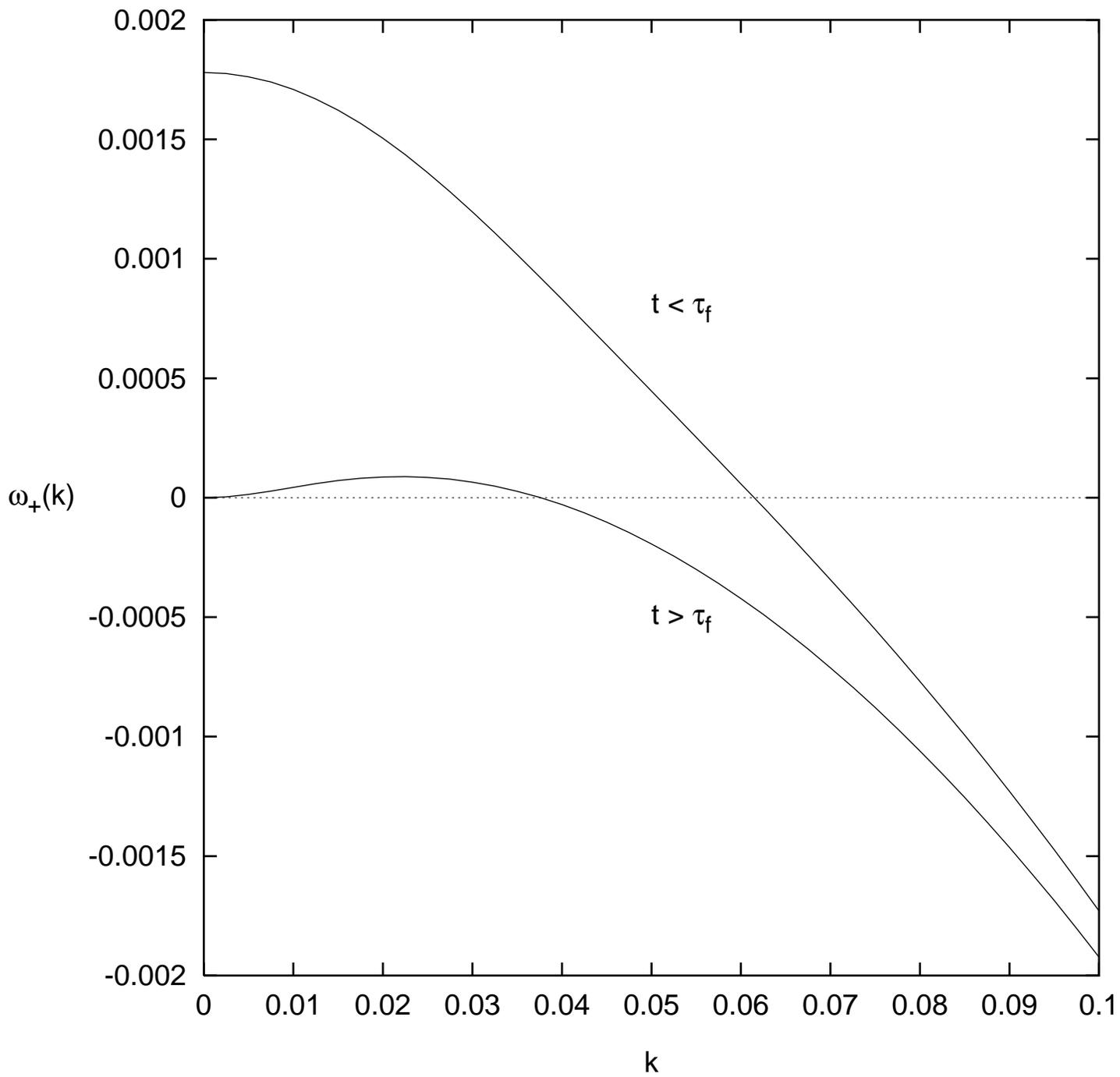
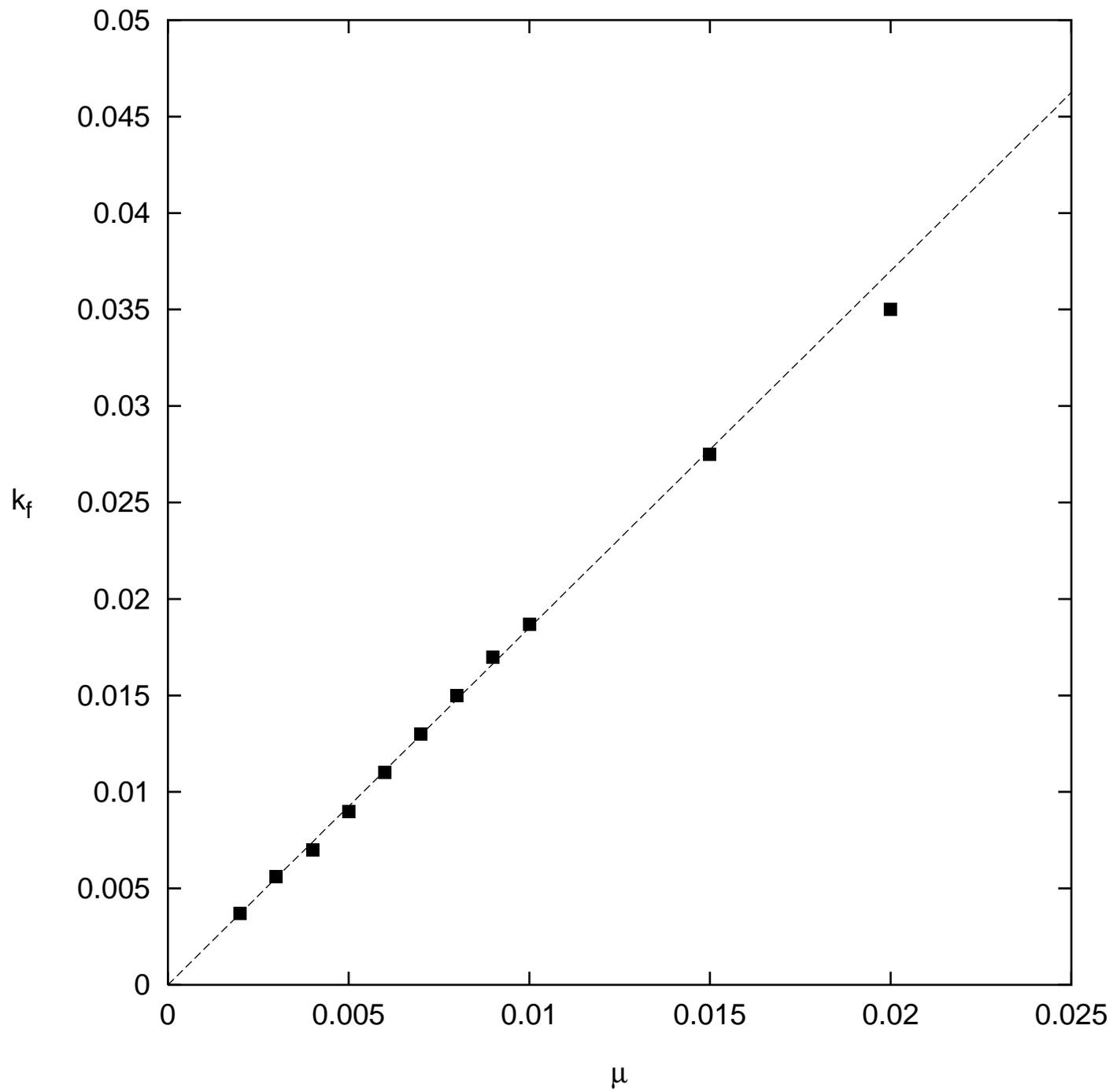
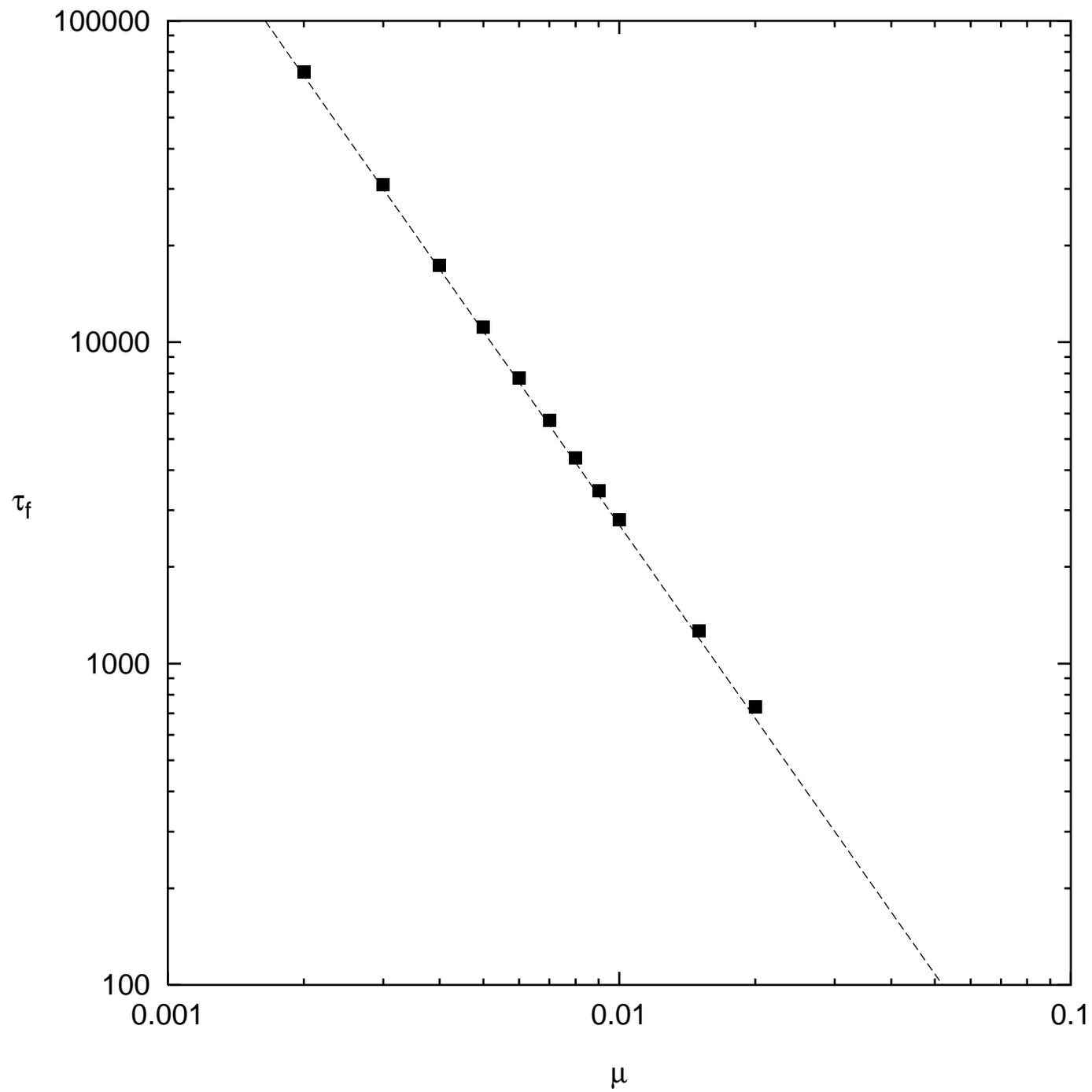
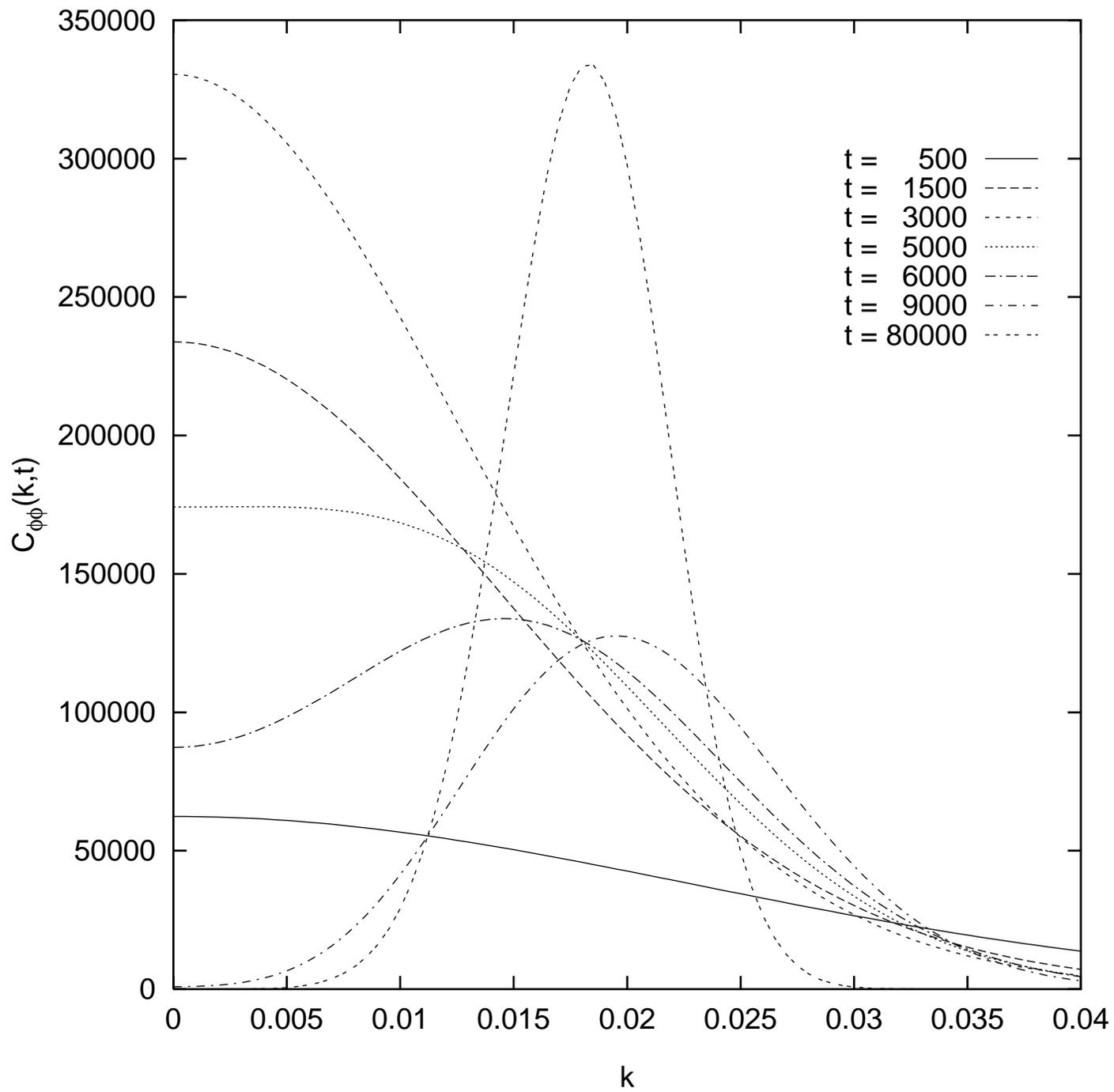
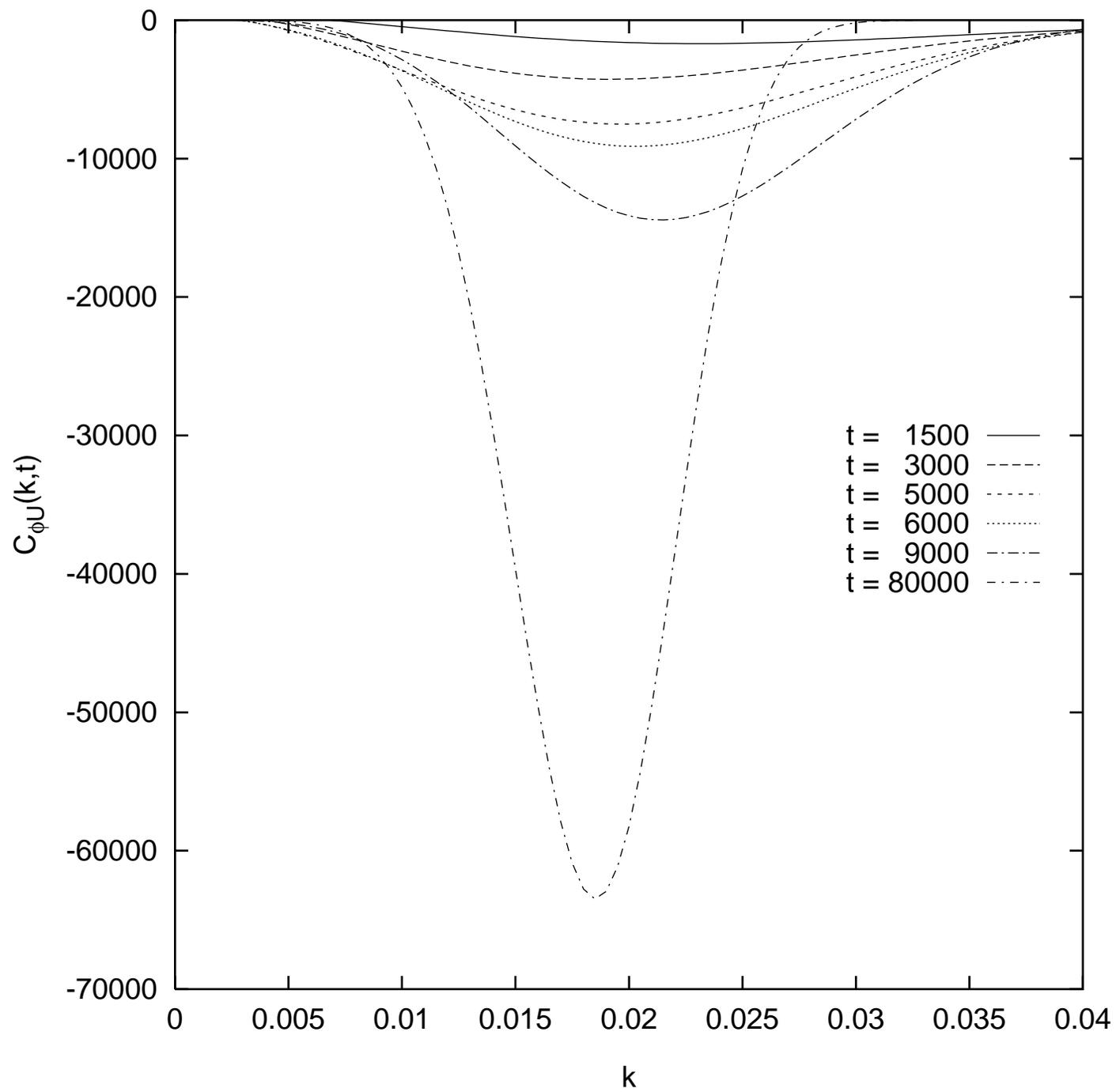
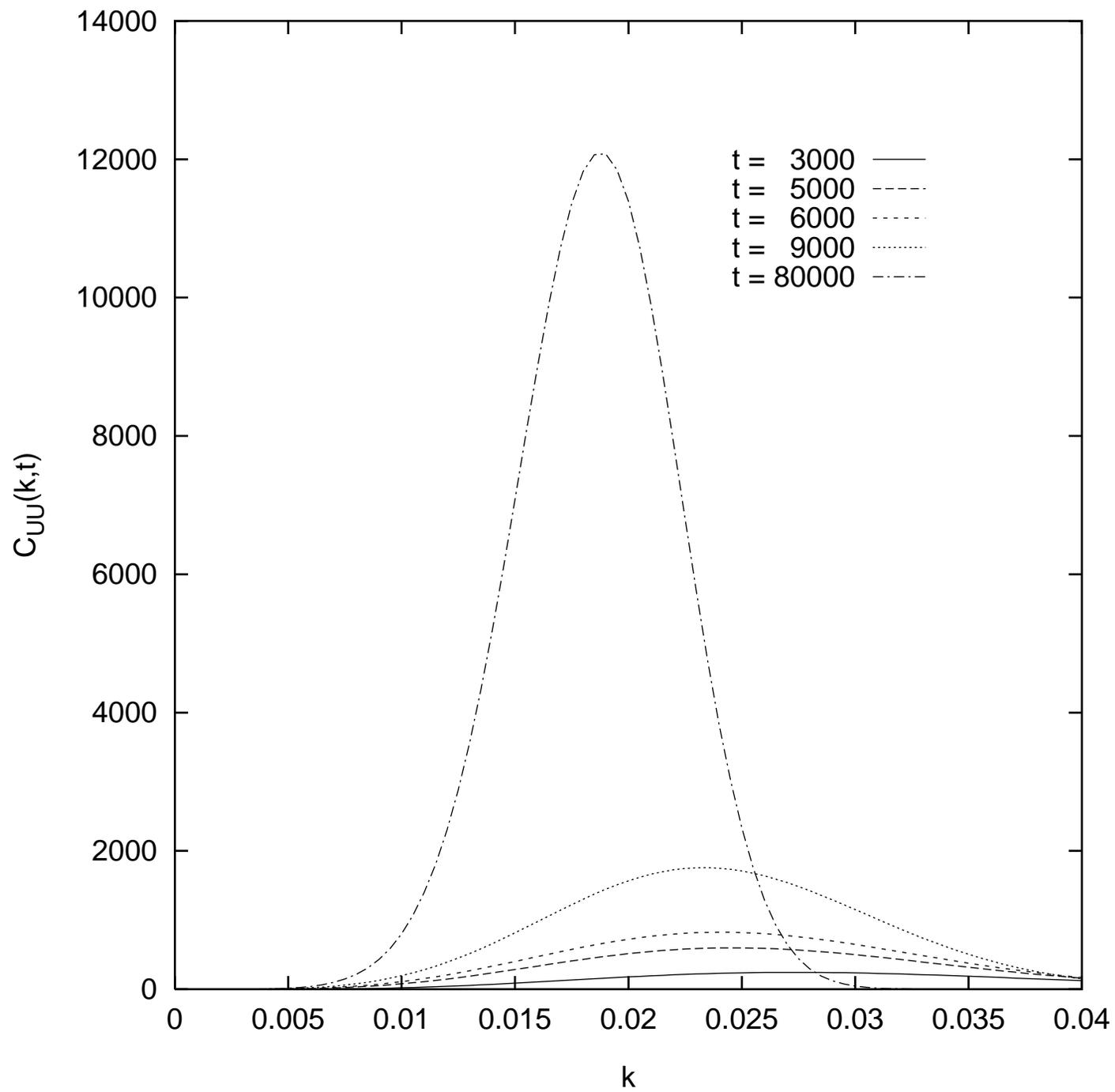
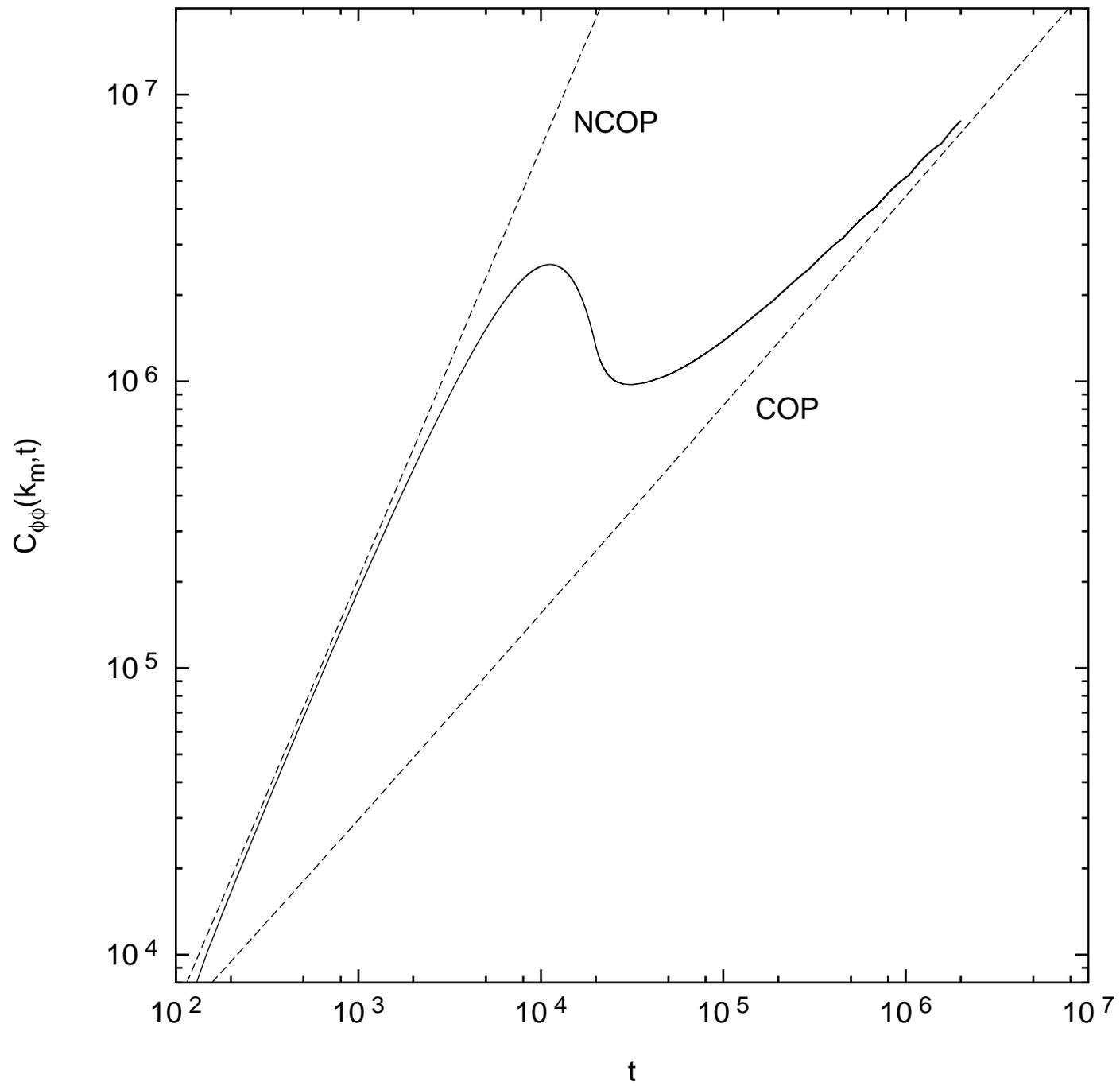
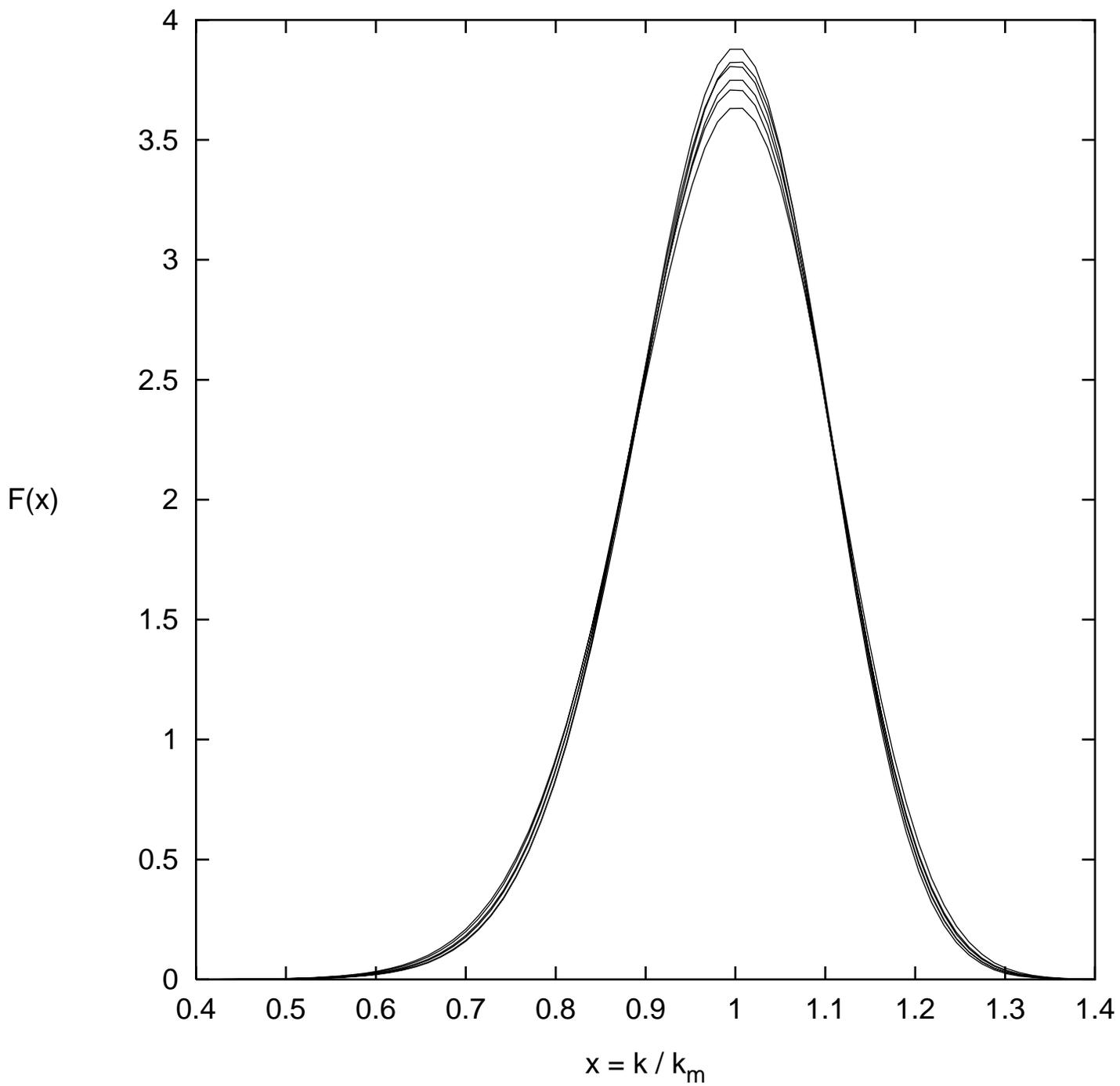
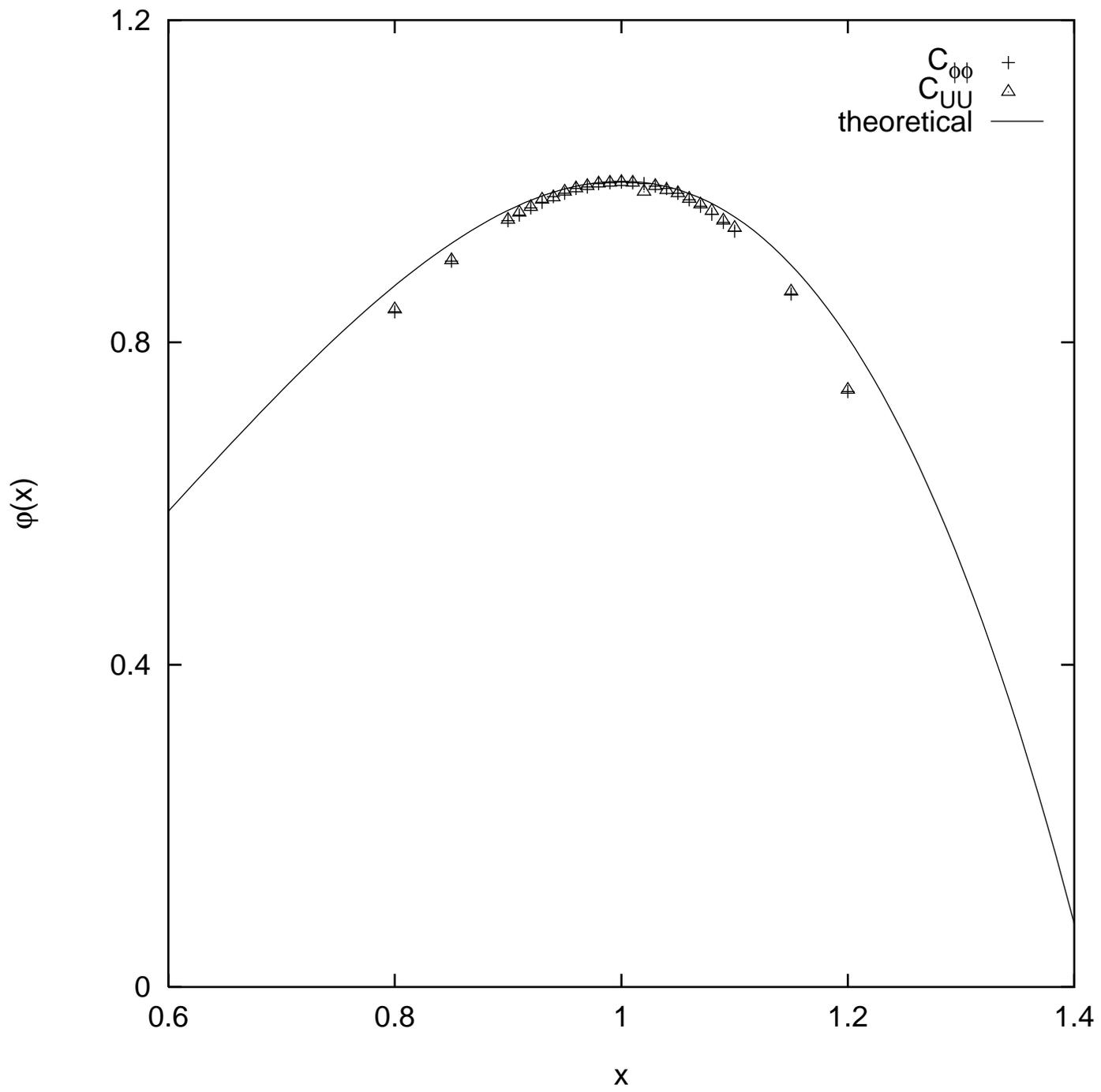
\end{document}